\documentclass{elsart3}

 \usepackage{graphicx}

\usepackage{amssymb}

\begin{document}

\begin{frontmatter}



\title{Ring Exchange and Correlated Fermions}


\author{Michel Roger}

\address{DSM-DRECAM, Service de Physique de l'Etat Condens\'e \\
Orme des Merisiers, CEA Saclay \\
91191 Gif sur Yvette Cedex. France}

\begin{abstract}
The concept of exchange in strongly-correlated fermions is reviewed with
emphasis on the generalization of the Heisenberg pair exchange to higher order
$n$-particle permutations. The ``frustration'' resulting from competing
ferromagnetic three-spin exchange and antiferromagnetic two- and four-spin
exchanges is illustrated on a two-dimensional model system: solid $^3$He films.
Recent experimental results proving the presence of significant four-spin
exchange interactions in the CuO$_2$ plaquettes of high $T_c$ cuprates are
reported.
\end{abstract}

\end{frontmatter}

\section{Historical introduction}
\label{}
The concept of exchange interactions in almost-localized correlated 
fermions first appeared in the pioneering papers of 
Heisenberg\cite{Heisenberg} and was formulated in a more general mathematical
way by Dirac\cite{Dirac1,Dirac2}. Although the early papers by Dirac already
contain the general expression of the Hamiltonian in terms of n-particle
permutations, no attention was paid, up to the sixties, to higher orders than
the ``pair-exchange'' Heisenberg term. Thouless\cite{Thouless}
 was the first to point out
that higher order exchanges as three- and four-spin cyclic permutations might
be important in quantum solids like $^3$He. But only ten years later, striking
experimental results on nuclear magnetism in the bcc phase of 
solid  $^3$He, in the millikelvin  range\cite{Adams,Godfrin,Osheroff},
 were interpreted by Hetherington, Delrieu and Roger through a ring-exchange
model with two- three- and four-particle interactions of 
comparable magnitude\cite{Hetherington,Delrieu,RDH}.
From general considerations put forward by Thouless, permutations of even parity
(like three-particle cycles) induce ferromagnetism while odd permutations
(pair and four-particle exchanges) favor antiferromagnetism, and the striking
phase diagram of bcc solid $^3$He corresponds 
to a highly frustrated quantum-spin system with 
competing three- and four-spin permutations. During the last
two decades a lot of progress has been accomplished in the investigation of
solid $^3$He films adsorbed on graphite, a simple model system exhibiting
even more frustration since  the frustrated nature of ring-exchange interactions 
is enhanced by the frustrated geometry of the triangular lattice.
The conceptual beauty of solid   $^3$He lies in the fact that the system is
simple enough (the pair interactions between  $^3$He atoms are mainly of hard
core nature) to allow the calculation of exchange frequencies from first
principles\cite{Ceperley} and a quantitative comparison with experimental
results.

The relevance of the multi-spin exchange concept is not restricted to the
physics of nuclear $^3$He spins. Delrieu\cite{unpub}
suggested that three-spin exchange might be dominant in the two-dimensional
electron Wigner solid near the quasi-classical limit, and this has been
corroborated through WKB calculations\cite{RogerWKB}. More recent Monte-Carlo
simulations have proved that competing three and four-spin exchange interaction
should occur near melting\cite{Bernu}. 

The first fourth-order $t/U$ expansion of 
the one-band Hubbard Hamiltonian in terms of four-spin interactions has been
published in 1977 by Takahashi\cite{Takahashi}. Soon after the discovery
of high-T$_{\rm c}$ superconductors, Roger and Delrieu\cite{RD} suggested,
on the basis of an expansion of the three-band Hubbard model, that
four-spin exchange  might be significant in the CuO$_2$ planes of cuprates.
During the last ten years many experimental results have revealed the
presence of four-spin exchange interactions in the Cu-O plaquettes of 
cuprates\cite{Sugai,Coldea,Adina} and copper-based spin-ladder
 materials\cite{Lorenzana,Imai,Eccleston,Mikeska}.
\section{Dirac formalism with illustration on the one-band Hubbard model.}
Dirac formalism is introduced in the framework of degenerate perturbation
theory.
The Hamiltonian is written 
$H=H_0 + V$,
 where $H_0$ describes
independent particles and V is a perturbation. 
The ``unperturbed'' degenerate ground-states for 
N distinguishable particles can be written as products
of independent particle states:
\begin{equation}
|\psi\rangle =|\alpha_1\rangle ^{(1)}|\alpha_2\rangle ^{(2)}
|\alpha_3\rangle ^{(3)}\cdots |\alpha_N\rangle ^{(N)}
\end{equation}
which means that the particle numbered (i) is in a state $|\alpha_i\rangle $.
Each of these states is itself a product of two kets 
corresponding to the orbital and spin variables respectively:
$|\alpha_i\rangle =|R_i\rangle |\sigma_i\rangle $
\begin{equation}
|\psi\rangle=|R_1\rangle ^{(1)}|\sigma_1\rangle ^{(1)}\cdots
|R_n\rangle ^{(N)}|\sigma_N\rangle ^{(N)}
\end{equation}
Hence $|\psi\rangle $=$|\phi^R\rangle |\xi^\sigma\rangle $ appears as a product
of an orbital wavefunction:
\begin{equation}
|\phi^R\rangle =|R_1\rangle ^{(1)}|R_2\rangle ^{(2)}\cdots |R_N\rangle ^{(N)}
\end{equation}
and a spin wavefunction:
\begin{equation}
|\xi^\sigma\rangle =
|\sigma_1\rangle ^{(1)}|\sigma_2\rangle ^{(2)}\cdots |\sigma_N\rangle ^{(N)}
\end{equation}
Note that for the half-filled Hubbard model, $|R_i\rangle $ simply represents 
the site occupied by the particle (i). The permutation P of two
particles can be expressed as a product of two operators:
\begin{equation}
P=P^R P^\sigma
\end{equation}
$P^R$ acting on orbitals and $P^\sigma$ acting on spin variables.
If the Hamiltonian does not depend explicitly on the spin,
 we can as a first step concentrate on the
orbital part of the wave function and solve the eigenvalue problem:
\begin{equation}
H|\phi^R\rangle =E|\phi^R\rangle 
\end{equation}
for the orbital wavefunction $|\phi^R\rangle $ describing
 N {\it distinguishable} particles.
The ground-state of the unperturbed part $H_0$ of the Hamiltonian is
N! fold degenerated and the corresponding subspace $\Omega_0$ 
is spanned by the N! states:
\begin{equation}
P^R |\phi^R\rangle  = |R_{\nu_1}\rangle ^{(1)}
|R_{\nu_2}\rangle ^{(2)}\cdots |R_{\nu_N}\rangle ^{(N)}
\end{equation}
where $\{\nu_1, \nu_2,\cdots,\nu_N\}$ represents a permutation $P$ of
the N integers  $\{1, 2,\cdots,N\}$.
We now apply degenerate perturbation theory\cite{Kato,Bloch} to the
perturbed Hamiltonian $H=H_0+V$. At first order, the splitting of the N!
degenerated energy levels is given by the eigenvalues of the Hamiltonian
$V^{(1)}$ defined by its matrix elements:
\begin{equation}
V_{a,b}^{(1)}=\langle\phi^R|P_a^RVP_b^R|\phi^R\rangle 
\end{equation}
where $P_a^R$ and $P_b^R$ are two permutations of the symmetric 
group ${ S}_N$. Since $V$ is invariant with respect to any  
permutation, we can write:
$$
V_{a,b}^{(1)}=\langle\phi^R|VP_a^RP_b^R|\phi^R\rangle =
\langle\phi^R|VP^R|\phi^R\rangle =V_P^{(1)}
$$
where $P=P_a^RP_b^R$, and the eigenvalue problem, restricted to the subspace 
$\Omega_0$ can be formally represented by the Hamiltonian:
\begin{equation}
H^{(1)}=-\sum_{P^R\in { S}_N} V_P^{(1)} P^R 
\end{equation}
where the summation runs over permutations $P^R$ of the symmetric
group ${ S}_N$. This result extends straightforwardly 
to higher order  degenerate
perturbation theory: the higher orders are expressed in terms 
powers of $V$ and projection operator $P_0$ on
$\Omega_0$\cite{Kato,Bloch}, and  these operators commute with 
permutation operators. Hence, at arbitrary order in degenerate perturbation
theory, we can write:
\begin{equation}
H\approx-\sum_{P^R\in { S}_N} V_P P^R 
\end{equation}

\noindent
We now have to introduce the spin degrees of freedom an express that the
global wavefunction is completely antisymmetric. As a general result from
Group theory a completely antisymmetric wave function can be expressed by the 
following bilinear expression\cite{Landau}:
\begin{equation}
|\psi\rangle =\sum_{\lambda,\mu} c_{\lambda,\mu}|\phi_\lambda^R\rangle
 |\xi_\mu^\sigma\rangle 
\end{equation}
where $|\phi_\lambda^R\rangle $ represents a linear combination of different
permutations $P^R_a|\phi^R\rangle $  corresponding
to a given irreducible representation of the symmetry group schematized by a 
Young diagram, while $|\xi_\mu^\sigma\rangle $ represents a linear
combination of permutation 
$P^\sigma_{\bar a}|\xi^\sigma\rangle $ corresponding to 
the 
representation associated with the ``complementary'' Young diagram, obtained
by exchanging the lines and the columns.
 For spin-1/2, a complete antisymmetrisation of the spins cannot
be realised over more than 2 variables, hence the corresponding Young
diagrams have at most two lines and each diagram corresponds to a given value
of the total spin $S$. It is then possible
to establish a correspondence between the expression [Eq. (10)] of the
Hamiltonian acting only on the orbital variables with an equivalent
Hamiltonian acting only on the spin variables.
Expressing the antisymmetry of the wave function:
\begin{equation}
P|\psi\rangle  = (-1)^p |\psi\rangle 
\end{equation}
where $p$ is the parity of the permutation, we can write:
\begin{equation}
P^RP^\sigma|\psi\rangle  = (-1)^p |\psi\rangle 
\end{equation}
and multiplying to the left by $(P^R)^{-1}$:
\begin{equation}
P^\sigma|\psi\rangle  = (-1)^p(P^R)^{-1} |\psi\rangle 
\end{equation}
Taking into account that in Eq. (10) $P^R$ and the inverse permutation
$(P^R)^{-1}$ appear with the same weight $V_P$, the Hamiltonian is
written equivalently in spin space:
\begin{equation}
H\approx-\sum_{P^\sigma\in { S}_N} (-1)^pV_P P^\sigma
\end{equation}
where the sum is over all permutations of the symmetric group ${ S}_N$
acting on spin variables.

Large U expansions of the one-band Hubbard model\cite{Takahashi} have 
generally been performed within the framework of second quantization
(i.e working on the total antisymmetric wavefunction). Dirac formalism
allows a simpler and more physical derivation.
We start with N distinguishable particles on N discrete
sites, each site containing zero, one or two particles. 
There is an onsite
 Coulomb energy $U$ for putting two particles on the same site, and
each particle has a probability $t$ to hop from one site to the nearest neighbor.
The Hamiltonian does not depend explicitly on the spin variables.

In perturbation theory, we start from the infinite $U$ limit: each site
is occupied by one particle. Since the particles are considered as
distinguishable,
the ground state is N! fold degenerated. We apply degenerate perturbation
theory to express the energy splitting as an expansion in $t/U$ when
$U$ is large but finite\cite{Kato,Bloch}. Up to fourth order, the effective
Hamiltonian restricted to the subspace $\Omega_0$ spanned by the
N! fold degenerated ground states is written\cite{Takahashi,Kato,Bloch}:
$$
{ P}_0VSV{ P}_0 + [{ P}_0VSVSVSV{ P}_0 -
{ P}_0VS^2V{ P}_0VSV{ P}_0]
$$
where ${ P}_0$ is the projection operator onto the subspace $\Omega_0$
and 
$
S=(1-{ P}_0)/(E_0-H_0).
$
\begin{figure}[btp]
\begin{center}\leavevmode
\includegraphics[width=0.8\linewidth]{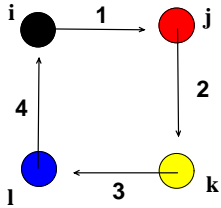}
\caption{ 
Evaluation of the four-particle cyclic permutation frequency. Only the
term ${ P}_0VSVSVSV{ P}_0$ in the fourth order expansion contribute
to the cyclic exchange of the four particle coordinates $i\to j$,
$j\to k$, $k\to l$ and $l\to i$. Each of these four particle hops is
accomplished by one of the four $V$ in the previous term and at each step
a factor $-t_{\alpha\beta}/(\nu U)$ where $\nu$ is the number of 
doubly occupied sites,
in the corresponding intermediate virtual state,  has to be
taken into account (this factor arises from the operator $S$). There are
4!=24 terms depending on the order in which the four hops 1, 2, 3, 4,
are successively realised. Eight of them  [corresponding to the sequences
(1324), (1342), (3124), (3142), (2413), (4213), (2431), (4231)] 
contribute to a factor $-t^4/(2U^3)$, 
the sixteen others contribute to
$-t^4/U^3$. The total contribution is then 
$-20t^4/U^3$.
}
\end{center}\end{figure}
This expression can be expanded in term of two, three and four-particle 
permutations $P^R$ acting on coordinates:
\begin{eqnarray}
 H=
 -J_1\sum_{<i,j>}^{(1)} P_{ij}^R -
 J_2\sum_{<i,k>}^{(2)} P_{ik}^R -
 J_3\sum_{<i,l>}^{(3)} P_{il}^R - &\nonumber \\
 J_T\sum_{<i,j,k>}[ P_{ijk}^R + ( P_{ijk}^R)^{-1}]- &\nonumber \\ 
 K  \sum_{<i,j,k,l>} [ P^R_{ijkl}+ (P^R_{ijkl})^{-1}] &\nonumber  
\end{eqnarray}
with:
$J_1=2(t^2/U)(1+4t^2/U)$,
$J_2=12t^4/U^3$,
$J_3=2t^4/U^3$, 
$J_T=10t^4/U^3$ and
$K=20t^4/U^3$ for the square lattice. 
The sums in the first line are on first, second and third
neighbor pairs respectively. The sum  run over triangles in the second
line and over square plaquettes in the last line.
As a typical example, the Fig. 1 illustrates the evaluation
of the four-particle cyclic permutation term. 

Since cyclic two and
four-particle permutations are of odd parity, while three-particle 
permutations are even, we obtain the corresponding  exchange Hamiltonian,
with permutation operators $P^\sigma$ acting on spin variables::
\begin{eqnarray}
 H_{ex}=
& J_1\sum_{<i,j>}^{(1)} P_{ij}^\sigma +
 J_2\sum_{<i,k>}^{(2)} P_{ik}^\sigma +
 J_3\sum_{<i,l>}^{(3)} P_{il}^\sigma - & \nonumber \\
& J_T\sum_{<i,j,k>}[ P_{ijk}^\sigma + ( P_{ijk}^\sigma)^{-1}]+ \nonumber \\ 
& K  \sum_{<i,j,k,l>} [ P^\sigma_{ijkl}+ (P^\sigma_{ijkl})^{-1}]  
\end{eqnarray}
Using the identity:
$
P_{i,j}^\sigma= (1+\sigma_i\cdot\sigma_j)/2
$
and the decomposition of any permutation in a product of transpositions,
it is easy to check that this Eq. 16  is identical to the result given by
Takahashi. However our previous expression clearly separate the ferromagnetic
contributions arising from three-particle exchange and the antiferromagnetic
ones coming from pair- and four-particle exchange. The formulation in terms
of permutation operators is also generally more convenient for the calculation
of the thermodynamic properties of a n-particle exchange Hamiltonian.
\section{Cyclic exchange in solid $^3$He and in  the Wigner Solid 
from first principles}
Dirac formalism has been generalized by Thouless for the exchange-problem
in solid $^3$He\cite{Thouless} and by Herring for many-electron systems
\cite{Herring}. This has been applied later to ab-initio calculations
of multi-spin exchange frequencies in solid $^3$He and in the two-dimensional
Wigner-Solid. A pioneering multidimensional WKB calculation of the exchange
frequencies has been applied to the two-dimensional Wigner solid near the
quasi-classical limit\cite{RogerWKB}. It has proved the conjecture by Delrieu
that three-particle exchange should dominate in that regime. 
Although, at physical
densities, solid $^3$He is far from the quasi-classical regime, the
same WKB approach was extrapolated with qualitative results concerning the
 hierarchy of various cyclic-exchange frequencies.

With the rapid increase of the performances of computers, ab-initio calculations
of ring-exchange frequencies through Path Integral Monte Carlo (PIMC)
 became possible.
In contrast to WKB evaluation they are relevant to the whole range of physical
densities, including solid phases near melting where quantum fluctuations are
large. The  exchange frequencies for bcc solid $^3$He\cite{Ceperley} near
melting, are in excellent agreement with the experimental thermodynamic 
properties, which can be inferred through exact high-temperature series
 expansions.

Although the long-range Coulomb potential is completely different from the
hard core potential between  Helium atoms, WKB calculations near the 
classical limit\cite{RogerWKB,Katano}  and PIMC in the whole 
density range\cite{Bernu} give striking analogies
between solid $^3$He films and the electron Wigner solid, on the same triangular
lattice. In both systems, three-particle exchange dominates near the 
quasi-classical limit, leading to ferromagnetism, while near melting,
antiferromagnetic two- and four-particle exchange compete with ferromagnetic
three-particle exchange leading to highly frustrated antiferromagnetic systems.  

Close to melting, both systems might be prototypes of two-dimensional ``spin
liquids''\cite{Misguich}
 with no long range order at T=0. Experiments on the magnetism
of the Wigner Solid are now at the very beginning\cite{Okamoto}, but 
 the nuclear magnetism of solid $^3$He films has now been extensively studied.
\section{Two-dimensional solid $^3$He: a model system with tunable frustration.}
Exfoliated graphite offers large (10x10 nm) flat crystallites with a strong
adsorption potential for He. Up to two solid He films can be adsorbed on this
substrate, within a wide density-range. The magnetism of 2D helium films
has been thoroughly studied, down to ~ 10-100 $\mu$K in:
\begin{enumerate}
\item
submonolayer  solid $^3$He films on graphite\cite{Fukuyama2}
\item
 solid $^3$He films on graphite coated by a high density 
$^3$He\cite{footnote,Greywall,Saunders,Fukuyama1,Bauerle} or
$^4$He\cite{Collin,Ishimoto2} layer or by a HD 
bilayer\cite{Ishimoto1}.

\end{enumerate}
The main difference between (i) and (ii) is that, due to a stronger adsorption
potential, the first layer is closer to the pure 2D case. At the same density 
exchange frequencies in the first layer are 10 times lower than in the second
layer where $^3$He atom permutations are  made easier through excursion in the 
third dimension. However the same general trends are observed:
\begin{itemize}
\item
A low-density  commensurate solid phase is stabilized by the graphite
 potential in case (i)
and by the periodic potential of the first layer in case (ii). 
Its susceptibility
has an antiferromagnetic character with a negative Curie-Weiss constant.
Both susceptibility and specific heat present anomalous deviations with
respect to the usual asymptotic behavior at high temperature
(Curie Weiss law for the susceptibility and T$^{-2}$ behavior for the specific 
heat). These features are due to the frustrated nature of the system and
can be quantitatively fit through the ring-exchange 
model\cite{Bauerle}. Down to 10~$\mu$K no  long-range
ordering has been detected, which seems to confirm the presence of a 
spin-liquid state. An ultra-low temperature specific-heat 
measurement\cite{Fukuyama1} and susceptibility 
measurements\cite{Collin,Ishimoto2}
put a higher limit of 10-100~$\mu$K on a possible gap in the excitation 
spectrum. Hence solid $^3$He films could be the first
experimental realization of a gapless spin liquid.
\item
At higher densities, an incommensurate solid phase is observed. The 
susceptibility has a positive Curie Weiss constant, but the unusual features
of the specific heat indicate that we have a ``frustrated ferromagnetic''
phase with relatively larger three-spin exchange but significant competing 
four-particle exchange\cite{Bauerle}. The  frustration is easily ``tunable'' in
the sense that when the density is further increased, three-spin exchange
dominates leading to a more conventional ferromagnet.
\end{itemize}
Some anomalies in the specific heat are present only in the commensurate
phases. A very interesting interpretation has been recently proposed by
H. Fukuyama\cite{Fukuyama2}, 
in terms of ``ground-state vacancies'' surrounded by magnetic polarons.
This exciting idea established a close link with strongly interacting
almost-localized electron systems close to a Mott-Hubbard transition
(e.g the physics of cuprates!).

\section{Four-spin exchange in high-T$_{\rm c}$ Cuprates}

In the insulating phases of Cuprates, the first experimental results in
contradiction with a pure Heisenberg model were the anomalously broad 
multi-magnon Raman spectrum\cite{Sugai} and the infrared optical absorption
by phonon assisted multimagnon excitations\cite{Perkins} 
in  La$_2$CuO$_4$. Both results can be understood with
the occurrence of four-spin exchange\cite{Sugai,Lorenzana}.
However the key experiments which really invalidates
the pure Heisenberg model is the inelastic neutron study of the
magnon spectrum near  the Brillouin zone\cite{Coldea}. It is in agreement
with the presence of ferromagnetic three-spin exchange terms $J_T$ as those
appearing in the expansion of the one-band Hubbard model (Eq. 16).
However, due to accidental cancellations occurring for the N\'eel
phase of the square lattice, the linear spin-wave spectrum is insensitive
to the presence of four-spin exchange operators\cite{Adina}. 
An excellent way to probe directly the presence of four-spin terms
is the study of spin-spin correlations in the paramagnetic phase.
These correlation have been recently measured through
diffuse magnetic scattering of polarized neutrons 
in La$_2$CuO$_4$ at $350~K < T < 500~K$\cite{Adina}. The results are compared
to predictions from high-temperature series expansions of the Hamiltonian.
They prove definitively the presence of four-spin interaction with a ratio
$K/J_1$ of about 20\%.

The same Cu-O plaquettes are present in copper based quasi-1D spin ladder
materials Sr$_14$Cu$_{24}$O$_{41}$ and La$_6$Ca$_8$Cu$_{24}$O$_{41}$, 
with approximately the same Cu-O bonds and distances as in La$_2$CuO$_{4}$.
The first neighbor pair exchanges along rungs and legs are expected to
be roughly equal $J_{rung}\approx J_{leg} =J$. In that case the theory
predict a gap $\Delta\approx 0.5J$ in the excitation spectrum. The experimental
value of the gap is substantially lower: 
$\Delta_{exp}\approx 0.35J$\cite{Imai,Eccleston}. A relatively small 
four-spin exchange $K/J \approx 20\%$ accounts for the decrease of the 
gap\cite{Mikeska}. This effect is related to the frustration due
to four-spin exchange, which increases the density of low lying eigenstates.

{\it
The concept of multi-particle exchange, pioneered by Dirac at the
beginning of the 20$th$ century, now appears as essential in  the
physics of strongly-correlated fermion systems which have raised considerable
interest during the last decades.}




\begin{thebibliography}{00}
\bibitem{footnote}
The density of the completed first $^3$He layer is so high that its
magnetism is negligible.
\end{thebibliography}
\end{document}